\pdfoutput=1

\documentclass[floatfix, reprint, superscriptaddress, amsmath, amssymb, aps, pra]{revtex4-2}
\usepackage{graphicx}
\usepackage{xcolor}

\usepackage{multibib}
\usepackage[pdftex=true,hyperindex=true,colorlinks=true,hidelinks=false]{hyperref}
\usepackage{physics}
\usepackage{float}
\makeatletter
\let\newfloat\newfloat@ltx
\makeatother
\usepackage{algorithm}
\usepackage{algpseudocode}
\usepackage{hyperref}

\usepackage{cleveref}
\crefname{figure}{fig.}{figs.}
\Crefname{figure}{Fig.}{Figs.}

\newcommand{\cnot}[1][cnot]{\textsc{#1}}

\begin{document}
\title{Optimization of deterministic photonic graph state generation via local operations}
\author{Sobhan Ghanbari}
\affiliation{Department of Physics, University of Toronto, 60 St George St., Toronto, ON, Canada}
\affiliation{Quantum Bridge Technologies Inc., 108 College St., Toronto, ON, Canada}
\author{Jie Lin}
\affiliation{Quantum Bridge Technologies Inc., 108 College St., Toronto, ON, Canada}
\affiliation{Department of Electrical and Computer Engineering, University of Toronto, 10 King’s College Road, Toronto, ON, Canada}
\author{Benjamin MacLellan}
\affiliation{University of Waterloo, Department of Physics \& Astronomy, 200 University Ave., Waterloo, ON, Canada}
\affiliation{Institute for Quantum Computing, 200 University Ave., Waterloo, ON, Canada}
\affiliation{Ki3 Photonics Technologies, 2547 Rue Sicard, Montreal, QC, Canada}
\author{Luc Robichaud}
\affiliation{Quantum Bridge Technologies Inc., 108 College St., Toronto, ON, Canada}
\affiliation{Department of Electrical and Computer Engineering, University of Toronto, 10 King’s College Road, Toronto, ON, Canada} 
\author{Piotr Roztocki}
\affiliation{Ki3 Photonics Technologies, 2547 Rue Sicard, Montreal, QC, Canada}
\author{Hoi-Kwong Lo}
\affiliation{Department of Physics, University of Toronto, 60 St George St., Toronto, ON, Canada}
\affiliation{Quantum Bridge Technologies Inc., 108 College St., Toronto, ON, Canada}
\affiliation{Department of Electrical and Computer Engineering, University of Toronto, 10 King’s College Road, Toronto, ON, Canada}

\begin{abstract}
Realizing photonic graph states, crucial in various quantum protocols, is challenging due to the absence of deterministic entangling gates in linear optics. To address this, emitter qubits have been leveraged to establish and transfer the entanglement to photons. We introduce an optimization method for such protocols based on the local Clifford equivalency of states and the graph theoretical correlations of the generation cost parameters. Employing this method, we achieve a 50\% reduction in use of the 2-qubit gates for generation of the arbitrary large repeater graph states and similar significant reductions in the total gate count for generation of random dense graphs.
\end{abstract}

\maketitle
\section{Introduction}
Graph states are a family of multi-qubit states where the entanglement structure between qubit nodes is characterized by the edges of a graph \cite{briegel_persistent_2001, hein_entanglement_2006-1}. Photonic graph states, with nodes representing photonic qubits, are essential resources for the realization of measurement-based \cite{raussendorf_one-way_2001,briegel_measurement-based_2009} and fusion-based \cite{bartolucci_fusion-based_2023} fault-tolerant quantum computing as well as various quantum communication protocols, including quantum repeaters \cite{zwerger_measurement-based_2012,zwerger_universal_2013,zwerger_measurement-based_2016,azuma_all-photonic_2015}. Photons are unique among qubit architectures as they are mobile qubits, a feature necessary for many of the mentioned applications, and are relatively inexpensive to create and manipulate, with long coherence times.

While graph states are efficient to describe in theory \cite{GottesmanThesis}, realizing them on photonic platforms is challenging due to the lack of photon-photon interactions in linear optics \cite{knill_scheme_2001}. 
The main approaches used to-date for indirect establishment of entanglement between photons, i.e., creating graph edges, are (i) probabilistic approaches utilizing a sequence of the so-called ``fusion gates'', based on photon interference, measurement, and post-selection \cite{browne_resource-efficient_2005, grice_arbitrarily_2011}, and (ii) deterministic approaches utilizing a set of interacting quantum emitters such as quantum dots or nitrogen-vacancy centers \cite{russo_photonic_2018} as ancillary qubits to induce entanglement between the photons as they are emitted \cite{schon_sequential_2005,lindner_proposal_2009}. Between the two approaches, the former faces an exponential scaling of resource overhead and requires indistinguishable photons, while the latter is experimentally challenging due to limitations on the coherent control and the entanglement of quantum emitters. %
\begin{figure}[b]
\centering
\includegraphics[width=\columnwidth]{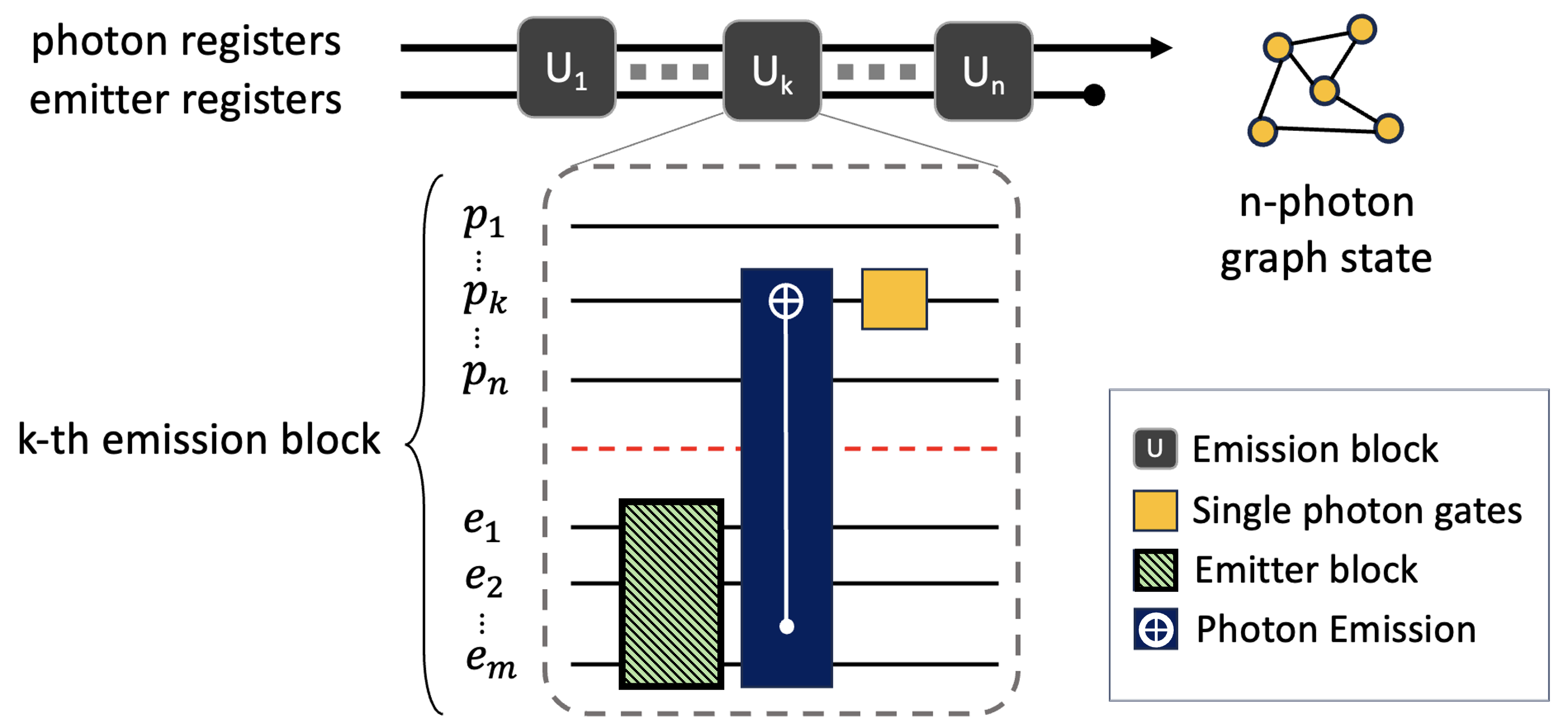}%
\caption{Schematic of the circuit used for the sequential generation of photonic graph states. The main components in each emission block are (i) single qubit gates on photons, (ii) an emitter block representing arbitrary operations on the set of emitter qubits between photon emissions, and (iii) a two-qubit emitter-photon gate corresponding to a photon emission event.}
\label{fig:seq_gen}
\end{figure}%
Because of better scalability and fueled by technological advancements, deterministic approaches have recently gained prominence in the community, as seen in experimental breakthroughs such as 10-qubit linear cluster states reported in quantum dot systems \cite{cogan_deterministic_2023} and 14-qubit GHZ states generated by a single memory atom in a cavity \cite{thomas_efficient_2022}. However, presently achievable graph sizes, still markedly limited by experimental factors \cite{schwartz_deterministic_2016, cogan_deterministic_2023,yang_sequential_2022}, remain far from the sizes required by many practical applications. Resource-efficient protocols are thus crucial for on-demand access to a broader, more practical range of graphs.

Deterministic methods use emitter qubits to generate, maintain, and distribute entanglement between photons as they are emitted. Figure \ref{fig:seq_gen} shows a quantum circuit representation for such protocols. The emission process, i.e. pumping an emitter followed by the emission of a photon, can be modeled as a 2-qubit gate, e.g., a \cnot{} \cite{lindner_proposal_2009}, between a non-existing photon (modeled by a qubit in $\ket{0}$) and an emitter, and is always the first gate applied on each photonic register. For systems designed to be deterministic, the main constraint in the model is not allowing multi-qubit gates on the emitted photons.

Devising the necessary operations between emission events for generating a target state is not trivial. As qubit connectivity is heavily constrained, this task differs fundamentally from general quantum circuit synthesis problems aimed at establishing a desired quantum channel \cite{hoyer_resources_2006,cabello_optimal_2011}. To date, only a limited number of resource-conscious generation protocols have been proposed, with the majority designed for specific categories of graphs \cite{lindner_proposal_2009,economou_optically_2010,buterakos_deterministic_2017,gimeno-segovia_deterministic_2019} and only a few for arbitrary states \cite{russo_generation_2019, li_photonic_2022, kaur_resource-efficient_2024}, with the optimization efforts concerned primarily with minimizing the number of emitters. Although the emitter count is a significant factor, there are other important metrics that can be considered in the circuit design, such as the required emitter coherence times and the number of two-qubit gates.

In this paper, we propose a method to reduce the generation costs of photonic graph states, by leveraging the concept of local Clifford (LC) equivalency of graphs \cite{van_den_nest_graphical_2004}. While LC equivalency has been previously employed to optimize the preparation of graph states \cite{cabello_optimal_2011, adcock_hard_2018, lee_graph-theoretical_2023}, this is the first time that it is applied to the case of deterministic photonic-state generation, differing notably from non-photonic cases. The method works through identifying a set of alternative intermediate graphs which can be converted to the final target state with minimal overhead {with respect to the photonic platform constraints}, providing a selection of generation routes to choose from. {While deterministic generation protocols such as Ref. \cite{li_photonic_2022} use Clifford circuits to generate a single {\it{fixed}} graph (the target), our approach differs by aiming to identify the best state to be generated, among {\it{all}} graphs in the local Clifford equivalence class. {\it{Any}} deterministic graph-to-circuit map can in principle be used to obtain the generation circuit for this process.} Moreover, we adopt a novel approach by employing graph theoretical metrics to correlate the structural characteristics of a graph with the cost parameters of its generating circuit, such as circuit size and depth. This is used to determine which intermediate graphs are more favorable for the process, filtered solely based on their shape. {The key insight of this work is leveraging the size and correlations of the LC equivalency class, leading to efficient and substantial reductions in generation costs. This lays the foundation for a versatile optimization framework for photonic state generation.}

{A primary result is the identification of the optimized elementary circuit blocks for generating arbitrarily large repeater graph states (RGS) \cite{azuma_all-photonic_2015}, achieving a 50\% reduction in required two-qubit gates between emitters. This reduction promises a potential orders-of-magnitude increase in the success rate of quantum repeater protocols. Moreover, we present for the first time a polynomial-time algorithm to generate the full local equivalency class of an RGS. We elaborate on the wide-range applicability of the new method by showcasing an efficient correlation-oriented optimization within the local equivalency classes of random dense graphs. We achieve average reductions of 65\% and 46\% in unitary gate count and `emitter depth' (defined in \Cref{background}), respectively, in generation circuits of 15-node graphs, and observe similar advantages as the graph size increases.}

\section{Background and preliminaries}
\label{background}
To introduce an unambiguous definition for the generation cost and the resources to optimize for, we consider the quantum circuit framework in \Cref{fig:seq_gen}. {The circuit models emitter qubits as entanglement mediators to produce a desired photonic state, taking a sequential block form with each block indicated by a photon emission event. Between successive emissions, arbitrary operations are allowed on the emitter space, but only single-qubit unitaries can be applied on the emitted photons. Such a configuration, with enough number of emitters, has been proved to be sufficient for on-demand generation of arbitrary graph states \cite{schon_sequential_2005}.} {To further simplify the cost analysis, the domain of operations considered in each step can be restricted to Clifford group unitaries and measurements in the computational basis, as they are proven to be universal for the preparation of graph states while being efficient to simulate classically \cite{GottesmanThesis,aaronson_improved_2004}.} 

{It should be noted that although the photonic gates can be applied in a sequential manner in each step, for each photonic register, one can combine all single qubit gates throughout the circuit into a single unitary to be applied at once. Therefore, the photonic part of the circuit has minimal contribution to its size. Additionally, the photonic gates are usually realized through linear optical elements, e.g., waveplates, and thus are easier to implement compared to the gates applied on the emitter qubits. Consequently, the emitter part of the circuit plays the primary role in cost considerations.}

{Notable circuit parameters to consider in a cost analysis are (i) the number of emitters, (ii) the number of gates of each type, and (iii) the circuit depth. In particular, using more emitters necessitates coherent control over a larger set of qubits, and a wider extent of emitter-emitter couplings, which is challenging in practice. This emitter coupling cost can be directly reflected in the number of two-qubit gates which, apart from being difficult to implement, are also a primary source of infidelity in the circuit. The depth of the circuit is associated with the maximum number of successive gates applied on the qubits and also indicates the time required to complete the protocol; two significant factors in circuit performance. As seen in protocols outlined in Refs. \cite{li_photonic_2022, buterakos_deterministic_2017}, an emitter qubit can undergo a reset after each measurement. Since the coherence time can be considered to reset as well in such cases, a more physically relevant metric would be the maximum number of consecutive gates applied on an emitter in the interval between each reset to \(\ket{0}\) and a subsequent measurement. Here, we define this new metric as `emitter depth', complementary to the regular circuit depth.

In our analysis, {while maintaining the minimum} emitter count, the focus is on other experimentally significant metrics that have mostly been overlooked in the existing literature on deterministic photonic state generation, e.g., the number of emitter-emitter \cnot{}s, and the emitter depth. Nevertheless, the optimization process can be easily adapted to incorporate other customized cost functions.}

\section{Optimization framework}
Our optimization method is based on two main notions: the local equivalency of states, and correlations between the structure of a graph, which is quantified by graph theoretical metrics, and the parameters of its generating circuit.

\begin{figure}[b]
\centering
\includegraphics[width=\columnwidth]{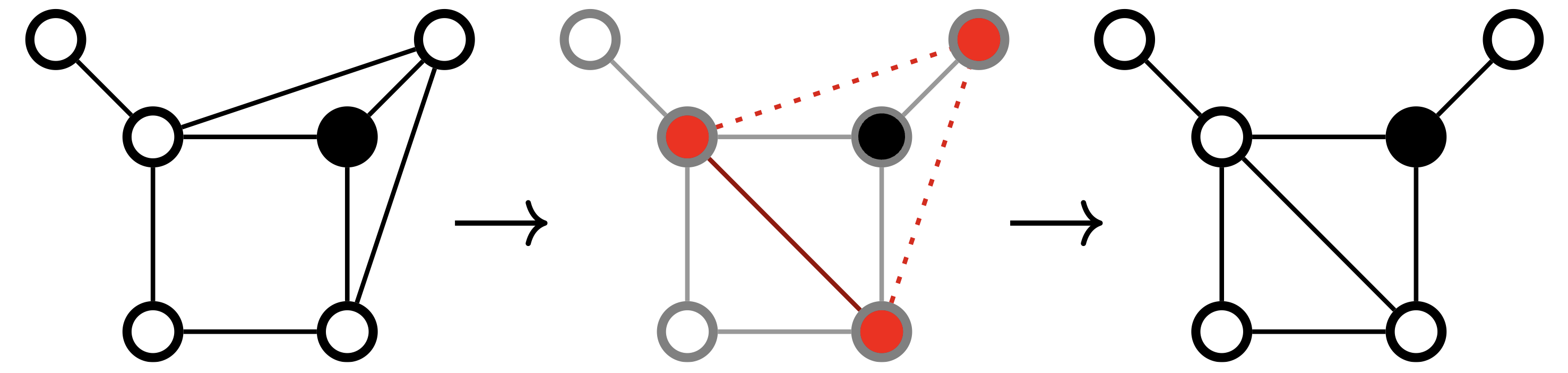}%
\caption{Example of graph transformation via local complementation on the black node in the initial graph on the left. The intermediate stage in the middle depicts the edge removal (dashed line) and edge creation (solid line) processes and the affected sub-graph denoted by the highlighted nodes.}
\label{fig:LC_intro}
\end{figure}

Two quantum states are locally equivalent if they can be converted into each other by local operations only, and are termed as locally Clifford (LC) equivalent if the operations are limited to single-qubit Clifford unitaries. It has been shown \cite{van_den_nest_graphical_2004, hein_multiparty_2004} that two graph states $\ket{G_1}$ and $\ket{G_2}$ are LC equivalent, if and only if their respective graphs can be transformed into each other through the successive application of a graph transformation rule known as local complementation (LC rule) \cite{bouchet_recognizing_1993}. When the LC rule is applied on node \textbf{\textit{a}} in a graph $G$, the subgraph comprised of the neighbourhood \textbf{{\(N_a\)}} of \textbf{\textit{a}} is complemented, while other edges in the graph remain intact (see \Cref{fig:LC_intro}). {This operation corresponds to the local unitary \cite{hein_multiparty_2004}:
\begin{equation}
	\begin{aligned}
     U_a = \sqrt{-i X_a}\prod_{b\in N_a}\sqrt{i Z_b}
	\end{aligned}
\end{equation}
in terms of Pauli operators  \(X_a\) and \(Z_b\) acting on the nodes \textbf{\textit{a}} and \textbf{\textit{b}}.}
Checking for the local equivalency of two graphs and finding a sequence of LC operations connecting them can be efficiently performed \cite{van_den_nest_efficient_2004}. Convertibility under local complementation forms a valid equivalence relation on the graphs. The equivalency class for any graph, also called an orbit, can be obtained by successive application of the LC rule on its nodes \cite{van_den_nest_graphical_2004, hein_multiparty_2004, adcock_mapping_2020}. 

In this paper, we will show that the LC orbit can be leveraged for cost reduction whenever a deterministic graph-to-circuit map, such as the protocol in Ref. \cite{li_photonic_2022}, is employed to find a generation circuit for a given graph. Since any two graphs within the same orbit can be converted to one another by local unitaries, i.e., single-photon gates, and given that\textendash as previously argued\textendash such gates do not significantly contribute to our circuit cost, the minimum generation cost for all orbit members is identical and is determined by the orbit member that yields the best circuit under the utilized graph-to-circuit mapping protocol. The goal is therefore to identify and generate a more favorable alternative graph in the orbit and then convert back to the original target.

Another set of alternative targets can also be obtained by permuting the temporal emission order of the photons, which provides a natural way of labelling the nodes in a graph. All isomorphic graphs found by changing the photon emission order share the same entanglement structure and thus can be considered identical once generated, conditioned by photon indistinguishability. The generation cost, however, varies from case to case as generating circuits depend on emission order \cite{li_photonic_2022}. 

{To compile an exhaustive array of alternative target graphs, the LC orbit is first identified via iterative application of the LC rule to graph nodes until no new graphs can be found. Subsequently, all possible emission orderings are considered for each orbit member.} {The set of alternative targets is then fed to a graph-to-circuit map in order to create a set of generating circuits, which are subsequently analyzed {to be ranked} with respect to a chosen {figure of merit.} {Our method takes the graph-to-circuit map as an input but operates at a higher level to expand the set of targets. While the proposed optimization scheme may utilize any mapping protocol, for concreteness, we have first focused on the protocol in Ref. \cite{li_photonic_2022} in our analyses.} This protocol guarantees the use of a minimal number of emitters for a given photon emission order, and runs in polynomial time, making it a suitable benchmark for evaluating cost reduction performance. Hereafter, for a given graph, comparisons are made between the original circuit obtained from \cite{li_photonic_2022}'s protocol and its optimized counterparts. {Lastly, to showcase that the framework can accommodate any graph-to-circuit mapping, we also implement the protocol from Ref. \cite{kaur_resource-efficient_2024} in the optimization, obtaining similar advantages to those achieved with Ref. \cite{li_photonic_2022}.}
}

\section{Results}
\begin{figure*}[t]
\centering
\includegraphics[width=\textwidth]{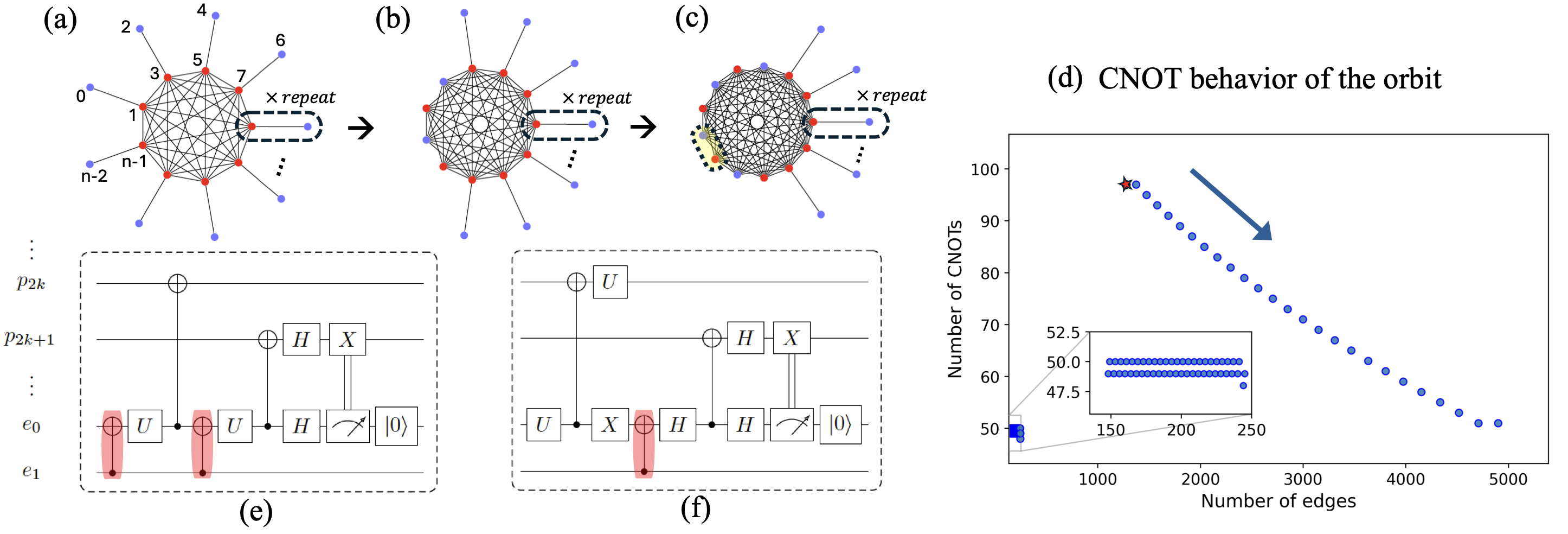}
\caption{The original RGS is shown in (a), along with partially transformed graphs in (b) and (c), corresponding to the correlated part of the \cnot{} behavior diagram shown in (d). In this diagram, the number of \cnot{}s is plotted against the number of edges for all the graphs in the orbit of a 100-qubit RGS, with the star-shaped marker indicating the original RGS. A linear correlation trend is visible within a portion of the orbit. The elementary circuit blocks in (e) and (f) are applied to the two emitters to generate the graphs shown in (b) or (c), adding the photon pair $p_{2k}$ and $p_{2k+1}$ to the state under generation. In these circuit blocks, emissions are modeled by the \cnot{}s between an emitter and a photon. From (e) to (f), the rate of use of emitter-emitter \cnot{}s (shaded in the circuit) changes from 1 per generated photon to 1/2. The circuit block in (e) generates a photon pair as in the repeated part of the original RGS structure, corresponding to a root and a leaf, while (f) generates a photon pair corresponding to the transformed part of the graph, indicated by the shaded pair of nodes in the graph (c). Gates $H$, $P$, and $X$ represent Hadamard, Phase, and Pauli-$X$, respectively, while $U$ is the Clifford unitary $HPHPX$. The final operations in both blocks are a measurement, a classically controlled gate based on the measurement result, followed by the emitter reset.}
\label{fig:RGS_pattern1}
\end{figure*}
{In general, finding the exhaustive set of alternative graphs is not always feasible except for small-sized or highly symmetric cases. This is because the LC orbit can grow rapidly with the size of the graph. For instance, for linear and cycle graphs the scaling is exponential \cite{bouchet_recognizing_1993}. Due to this complexity, previous investigations into the LC orbit have been limited to graphs with a maximum of 12 nodes \cite{cabello_optimal_2011, adcock_mapping_2020}. {However, identifying every member of the orbit is often unnecessary for gaining an advantage with respect to a given cost metric. In the following sections, we see cases where optimization can be achieved by recognizing certain patterns or correlations that allow us to find better performing graphs in a scalable and efficient manner without full orbit exploration.}
\subsection{Repeater graph states}
{For structured graphs that can be parameterized by size—{i.e., graphs that preserve a characteristic node connectivity structure across different sizes}—analyzing the full orbit of smaller instances can reveal general patterns applicable to larger graphs of the same type. The RGS, depicted in \Cref{fig:RGS_pattern1}(a), is one such structured graph, facilitating all-photonic quantum repeater protocols and enabling long-distance quantum communication \cite{buterakos_deterministic_2017,pant_rate-distance_2017,hilaire_resource_2021}. Exploring every member of the orbit of small RGSs indicates that the number of emitter-emitter \cnot{} gates can be reduced by nearly 50\%. This potential reduction can be seen in \Cref{fig:RGS_pattern1}(d) where we plot a correlation diagram for the circuit \cnot{} count as a function of the number of the edges in their respective graphs for the orbit of a 100-qubit RGS. We observe that the orbit can be divided into a linearly correlated region and an uncorrelated one (shown in the inset) with respect to the edge count. {By studying the correlated portion of the orbit, one can find the sequence of LC operations required so that the graph, and the corresponding generating circuit, move one step toward the lower-cost region as indicated by the arrow in \Cref{fig:RGS_pattern1}(d).} Figure \ref{fig:RGS_pattern1} (b) and (c) correspond to the first two graphs belonging to the linearly correlated part. To obtain the graphs shown in (b) and (c), local complementation operations (LC rule) are applied on the nodes of the original RGS shown in \Cref{fig:RGS_pattern1} (a) in order of $\{1, 3, 1\}$ and $\{1, 3, 5, 7, 1\}$, respectively. Here, the node labeling (starting from $0$) corresponds to the photon emission order as well. Similarly, to obtain the $j$-$th$ graph along the linearly correlated part ($j=0$ being the original RGS) we must apply the LC rule on the odd numbered nodes up to $4j-1$, followed by an additional LC rule on node $1$. Remarkably, breaking such an LC rule sequence at any node, results in a unique graph in the local equivalency orbit. By comparing the sequences leading to two consecutive graphs ($j$ and $j+1$) we observe that the $j+1$-$th$ graph has two extra nodes, $4j+1$ and $4j+3$, before the final LC on node 1 in its sequence compared to the $j$-$th$ one. Stopping the transformation sequence at each of these two points results in two graphs belonging to the uncorrelated region of the orbit respectively, constituting the inset of \Cref{fig:RGS_pattern1}(d).
 \begin{figure*}[]
\centering
\includegraphics[width=0.9\textwidth]{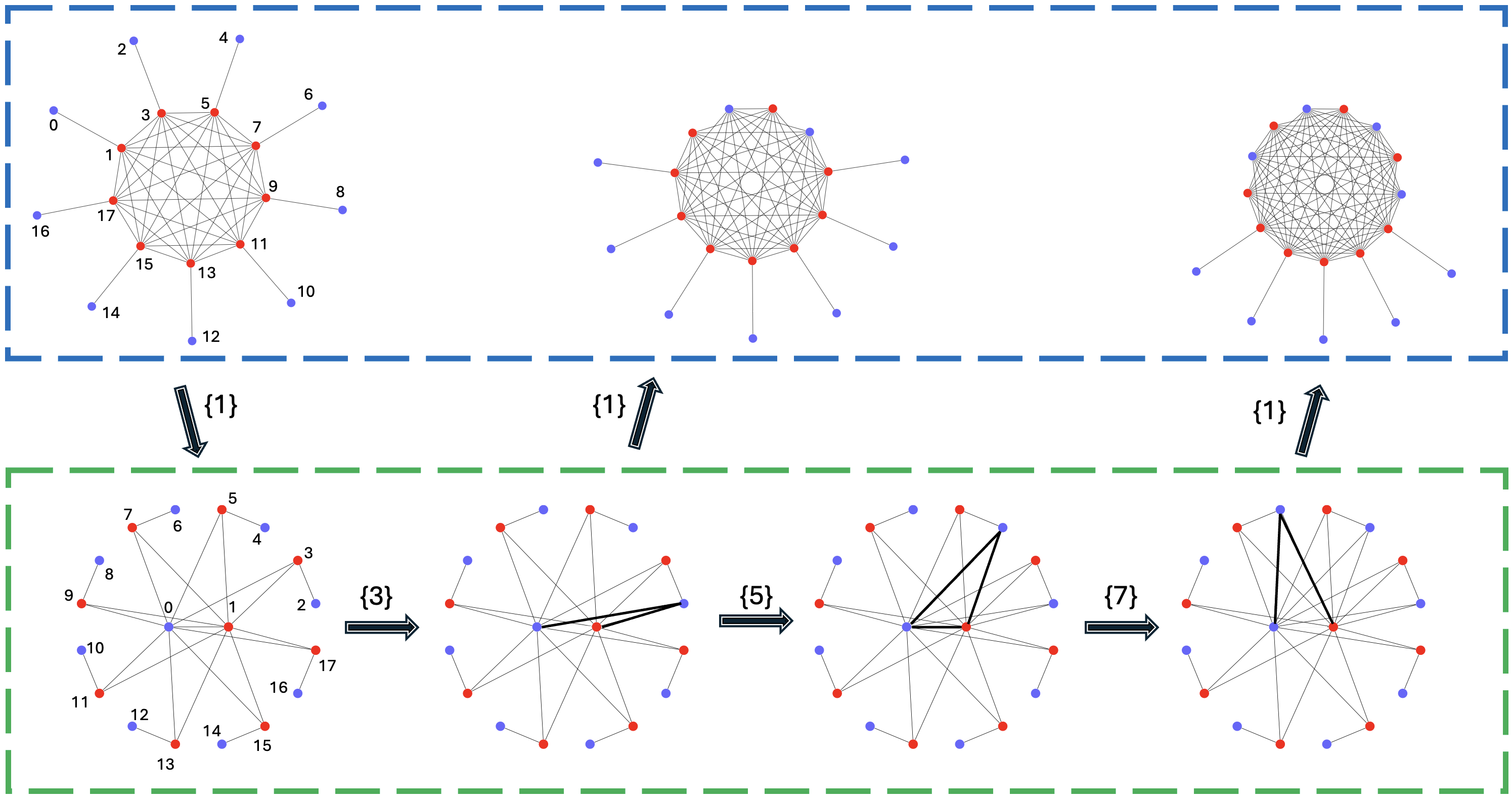}
\caption{Seven first graphs in the orbit of a 18-qubit RGS. The numbers on arrows indicate on what node the local complementation is applied to obtain the next graph. The observed pattern leads to the identification of an algorithm for a full orbit enumeration and being able to find elementary circuit blocks to realize each different shape. On top, from left to right, in each step a pair of arms from the RGS is folded into the central highly connected component. On bottom, from left to right, in each step one of the leaf nodes (even numbered) gets connected to nodes 0 and 1, and the nodes 0 and 1 alternate between being connected and not connected. The new edges created in each step are represented with bold lines.
}%
\label{fig:RGS_pattern}
\end{figure*}
\begin{figure}[h]
\centering%
\includegraphics[width=0.95\columnwidth]{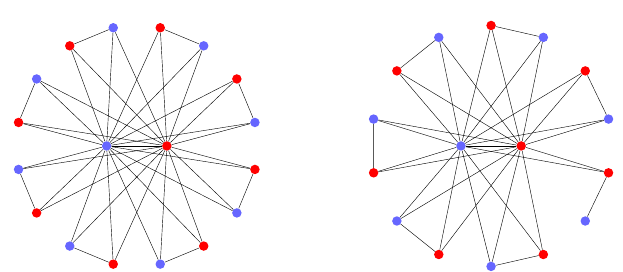}
\label{subfig:windmill_like_and_windmill}
\caption{The best alternative graph shapes in the orbit of an $n$-node RGS with odd (left) and even (right) number of arms ($n/2$= 9 and 8 respectively) in its original form. Larger cases can be obtained by adding pairs of nodes to the graphs such that the added two nodes are connected to each other and to the two central nodes.}
\label{fig:best_shapes}
\end{figure}
Figure \ref{fig:RGS_pattern} shows the pattern observed in the orbit and the required local complementation operations to navigate between them. The graphs in the top row correspond to the first few cases in the linearly correlated part of Fig. 3(d). The graphs on bottom represent the uncorrelated part of Fig. 3(d) (the inset).  

\begin{algorithm}
\caption{RGS orbit enumeration}
\label{alg}
\begin{algorithmic}
\State Define orbit as an empty list
\State Add the original RGS $G_0$ to the orbit
\State Define core-list as the list of indices of the central nodes (fully connected portion) in graph $G_0$
\State Apply LC transformation (local complementation) on the first node in the core-list of $G_0$ to get a new graph $G$. Refer to this node as $i$ hereafter
\State Add $G$ to orbit 
\State Remove $i$ from the core-list
\While{core-list is \textsc{NOT} empty}
    \State In $G$, apply LC transformation on the next first node $j$ in the core-list to get the new graph; set it as $G_1$
    \State Add $G_1$ to orbit 
    \State Remove $j$ from core-list
    \State In $G_1$, apply LC transformation on the core node $i$ to get the new graph; set it as $G_2$
    \State Add $G_2$ to orbit 
    \If{core-list NOT empty}
        \State In $G_1$, apply LC transformation on the next first node $k$ in the core-list to get the new graph; set it as $G_3$
        \State Add $G_3$ to orbit 
        \State Remove $k$ from core-list
        \State Set $G\gets G_3$
    \EndIf
\EndWhile
\end{algorithmic}
\end{algorithm}

\begin{figure*}[bt]
\centering%
\hspace{-1.75em}\includegraphics[scale=0.375]{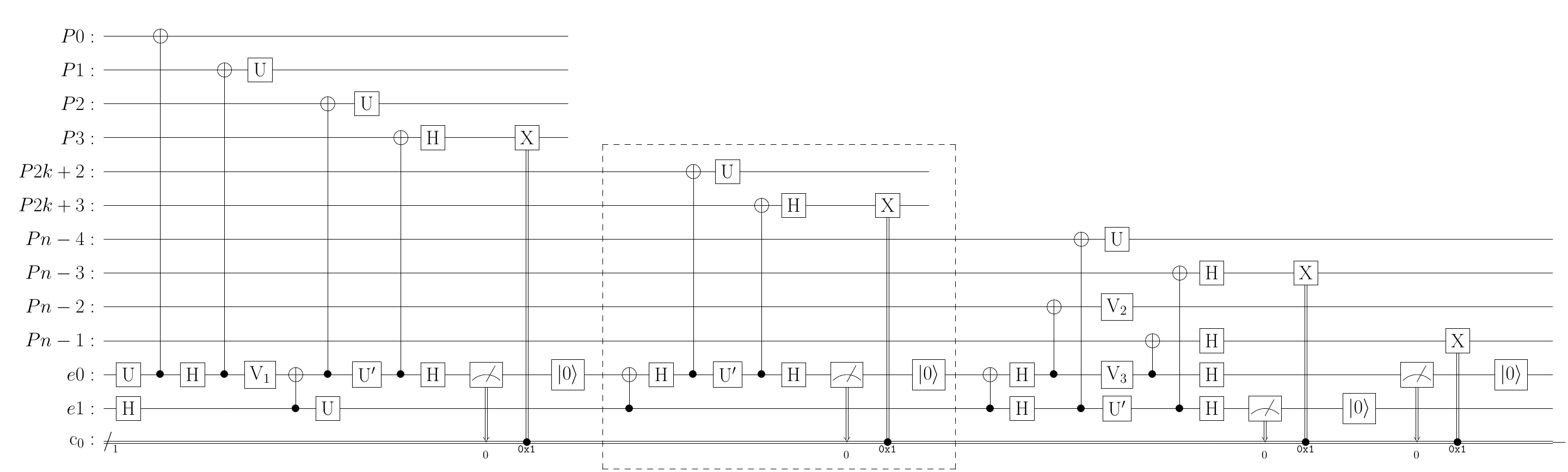}
\caption{The circuit to generate the best alternative graph in the orbit of an RGS with $n=2m+8$ qubits. The elementary circuit block in the middle is repeated for $k$ time with $1\leq k\leq m$ while the rest of the circuit is fixed. $U$ and $U'$ are Clifford unitaries $HPHPX$ and $HPHP$ respectively. For odd $m$: $V_1$ = X, $V_2$ = U, and $V_3$ = $U'$, and for even $m$: $V_1$ = I, $V_2$ = $V_3$ = H.}
\label{fig:windmill_circ}
\end{figure*}

Such generalization allows us to navigate the orbit to reach the graph with the minimum \cnot{} count. This also provides, for the first time, an efficient polynomial-time algorithm to generate and access the entire orbit of an arbitrarily large RGS. {The procedure is outlined in detail in \Cref{alg} below.} It is noteworthy that the LC orbit exploration via brute-force search methods is not an efficient and scalable process in general, {as one needs to perform a search over all possible sequences of local complementation operations on graph nodes and repeat the whole process for every new graph found in each round. In fact, even determining the size of the orbit for an arbitrary graph is a \#P-complete problem \cite{dahlberg_counting_2020}.} 

By looking into the generating circuit for each of the graphs in the orbit, one can find different elementary circuit blocks that are to be utilized to add photons to the state in a sequential manner, constituting different parts of the target structure. The generation circuit blocks in \Cref{fig:RGS_pattern1} (d-e) demonstrate how the discussed graph transformations can change the scaling of the required number of 2-qubit gates from $n$ to $n/2$ for an $n$-qubit RGS, since the emission of each photon pair belonging to the transformed part of the graph requires a single \cnot{} gate as opposed to the two gates needed when using the original RGS shape. Ultimately, the best (with respect to the use of \cnot{} gates) alternative graph in the orbit can be identified by continuing this observed trend, and is achieved by sequentially applying the LC rule to all central nodes of an RGS, i.e., LC rule on nodes $\{1, 3, \ldots, n-1\}$, if the number of arms ($n/2$) in the RGS is an odd number. For even $n/2$, the last node in the sequence is replaced by the leaf node connected to it, resulting in the set $\{1, 3, \ldots, n-2\}$. The \cnot{} count is, as a result, reduced from $n-3$ to $n/2-2$.} {The optimal cases are depicted in \Cref{fig:best_shapes} and their corresponding circuits are shown in \Cref{fig:windmill_circ}.} This optimized scaling, identified systematically here, aligns with the scaling of the ad-hoc protocol described in Ref. \cite{buterakos_deterministic_2017}, which was specifically designed for the optimal generation of RGS using a different approach and with another emission sequence. {Note that the identified pattern and the optimization result are subject to change when different graph-to-circuit mapping protocol are employed. In \hyperref[App:A]{Appendix A}, we expand on how using another graph state generation protocol outlined in Ref. \cite{kaur_resource-efficient_2024} results in scaling advantages in terms of circuit depth for RGS generation.}

{An important benefit of reducing the number of \cnot{} gates, aside from mitigating the technical complications of realizing the generation circuit, is that it nearly doubles the maximum size of the producible RGS. This is because the size of the state is primarily limited by the emitter's coherence time and how many emission events can be performed before the emitter qubits lose coherence, and the main time consuming operations in the circuit are the emitter-emitter \cnot{}s \cite{kaur_resource-efficient_2024}. The significance of this size increase is better reflected in the repeater protocol's success probability in establishing an EPR pair between two distant parties using multiple repeater stations in between, which is given by $P=(1-(1-P_b)^{n/4})^N$ \cite{azuma_all-photonic_2015}. Here, $n$ is number of photons in the used RGS, $N$ is the number of RGS stations between the two parties, and $P_b$ is the probability of performing a successful entanglement swapping between two photons coming from two adjacent repeater stations. Doubling the size from $n$ to $2n$, especially in the realistic long distance regime with $n$ being in the order of tens and $N$ in hundreds, results in several orders-of-magnitude increase in the overall success rate. See \hyperref[App:B]{Appendix B} for more details on how the optimization affects the size and performance of the RGS.}
\begin{figure}[b]
\centering
\includegraphics[width=\columnwidth]{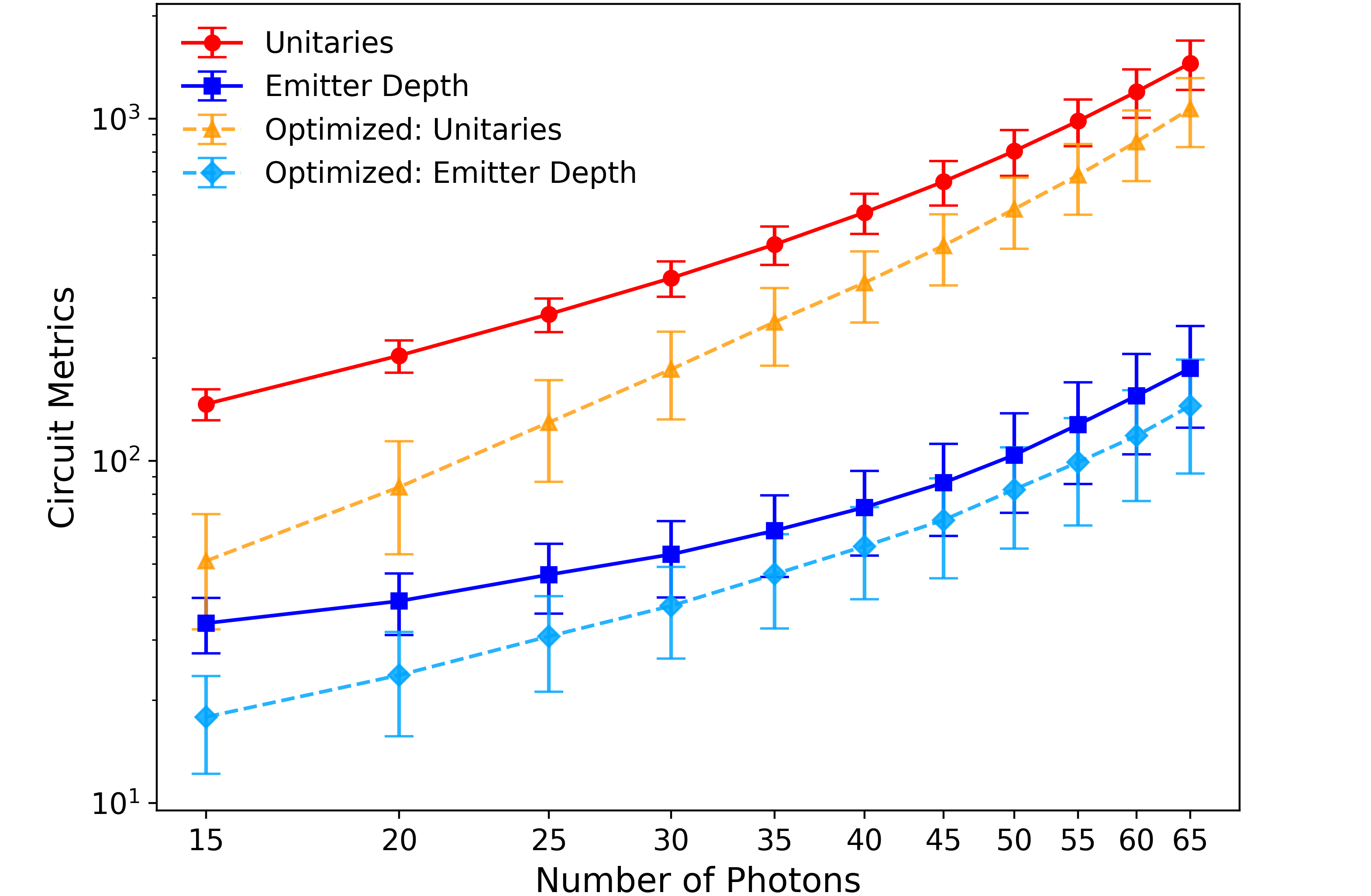}
\caption{Original (solid) and optimized (dashed) cost parameters for different graph sizes. Error bars are one standard deviation for values averaged over 1000 random graphs.
}
\label{fig:cost_reduction}
\end{figure}
\subsection{Non-structured graphs}
{To demonstrate the applicability of our method beyond structured graphs, we also investigated the opposite end of the spectrum, i.e., random graphs \cite{Erdos2022OnRG}. {In order to expect a generalizable optimization behavior, a shared trait is needed in the set of graphs under analysis. To this end, we limit the study to random dense cases, specifically focusing on graphs with a 95\% edge probability between each pair of nodes.} Conjecturing a correlation between the generation cost and the shape of a graph, we chose a simple metric to parameterize the graph shape: the number of edges. The existence of such a correlation can be directly confirmed for graphs of small size, e.g., with 15 nodes, as the whole orbit is accessible by exhaustive search. By choosing the graph with the least edge count in the orbit, we observed average reductions of $65\%\pm11\%$ in the total number of unitary gates of the circuit and $46\%\pm18\%$ in emitter depth. One can efficiently test this correlation for larger graphs, as there are efficient methods to find sparse graphs within the orbit of dense graphs; see \hyperref[App:C]{Appendix C} for one such efficient strategy developed by the authors, with no claim of being optimal. Figure \ref{fig:cost_reduction} shows the cost reduction achieved when extending the use of this strategy for graph of up to 65 nodes.}

{We also conducted an exhaustive analysis (see \hyperref[App:D]{Appendix D}) of graphs with up to 7 qubits \cite{dataset}, enumerating all the possible 45 entanglement classes as defined in Ref. \cite{hein_multiparty_2004} and comparing the generation circuits across $\sim1.9 \times 10^6$ labeled connected graphs \cite{numberofgraphs}. 

All the presented analyses were conducted on a 3.2 GHz, 8-core, Apple M1 Pro chip with 16 GB of RAM, using our Python package ``GraphiQ'' \cite{GraphiQ}.}

\section{Discussion}
We introduce an optimization framework tailored for the deterministic generation of \textit{photonic} graph states, a task that poses significant theoretical and experimental challenges due to its inherent constraints. Our approach leverages the concept of local Clifford equivalency and local complementation to effectively identify alternative graphs in the same local equivalency class that can serve as more resource-efficient paths for reaching a target state. 

In our correlation analyses, we establish that specific graph metrics can be linked to circuit cost parameters, enabling us to manage the use of resources at an orbit search level. {Future work should explore other graph structures of interest and with more elaborate graph theoretical metrics, to discover the optimal entanglement generation path.} 

It is worth noting that the employed cost metrics can be tailored to meet the requirements of specific experimental platforms, e.g., as we have done with the `emitter depth' to address the limitations in emitters' coherence time. Additionally, the application can be easily extended to the probabilistic fusion-based graph generation protocols, such as the one in Ref. \cite{bartolucci_fusion-based_2023}, by facilitating the expansion of both the size and diversity of resource states that are required as building blocks in making larger graphs.

\section*{Acknowledgement}
We thank Sophia Economou for helpful discussions. This work was supported in part by MITACS, and the Air Force Office of Scientific Research (AFOSR) under Grant FA9550-22-1-0062. Benjamin MacLellan acknowledges support from the NSERC Vanier CGS program. Hoi-Kwong Lo acknowledges research support from NSERC, Innovative Solutions Canada, and NRC (National Research Council of Canada) CSTIP Contribution Agreement QSP 081-1.

\section*{Appendix A: Graph-to-circuit maps}
\label{App:A}
In general, the optimization outcome is dependent on the specific deterministic graph to circuit map that is used to find a generating circuit for each input graph. However, the framework can take any such graph-to-circuit mapping protocol to work with. The first protocol we implemented (Ref. \cite{li_photonic_2022}) was selected to ensure that the algorithms utilized the minimum number of emitters necessary to generate the entanglement present in a graph state. However, the number of emitters may not always be the primary cost metric and thus other maps can become more suitable options depending on the platform and hardware. For instance, Kaur et al. \cite{kaur_resource-efficient_2024} have shown that considering solely the photon loss as the dominant factor in the failure of an all-photonic repeater protocol, a map that allows for the use of extra emitters (over the theoretical minimum) actually results in a better overall success rate. 

Since the optimization framework proposed here is applicable to other protocols, we analyzed repeater graph states of varying sizes using another specific algorithm (Algorithm 2, as referenced in \cite{kaur_resource-efficient_2024}). In this analysis, we employed four emitters instead of the minimum two required for RGS generation. The results are promising, revealing patterns analogous to those observed previously. For example, when plotting the \cnot{}-vs-edge count diagram (\Cref{fig:kaur_correlation}), two distinct parts of the orbit become recognizable, with one showing a strong linear correlation. It is worth noting that the strength of the employed algorithm is in its effectiveness in depth reduction, thus the number of \cnot{}s, even in the best case (scaling as $3\times\lfloor n/3 \rfloor$ with $n$ being the size of the graph), is larger compared to our previously identified optimal case ($n/2-2$). However, when considering circuit depth (to be specific emitter depth as defined in the main text) the results show an advantage over the circuit depths obtainable with the previous protocol ($n/2+6$ vs $n$ scaling). For the four-emitter scenario, the optimal graph featuring the least depth coincides with the case that requires the fewest \cnot{} gates. Additionally, the results can be straightforwardly generalized to arbitrary large RGS as we did before; no matter the size of the RGS, when using the orbit enumeration Algorithm (\ref{alg}), the optimal shape resulting in the above-mentioned advantages is always the third graph in the list of graphs added to the orbit via the algorithm.

\begin{figure}
    \centering
    \includegraphics[width=\columnwidth]{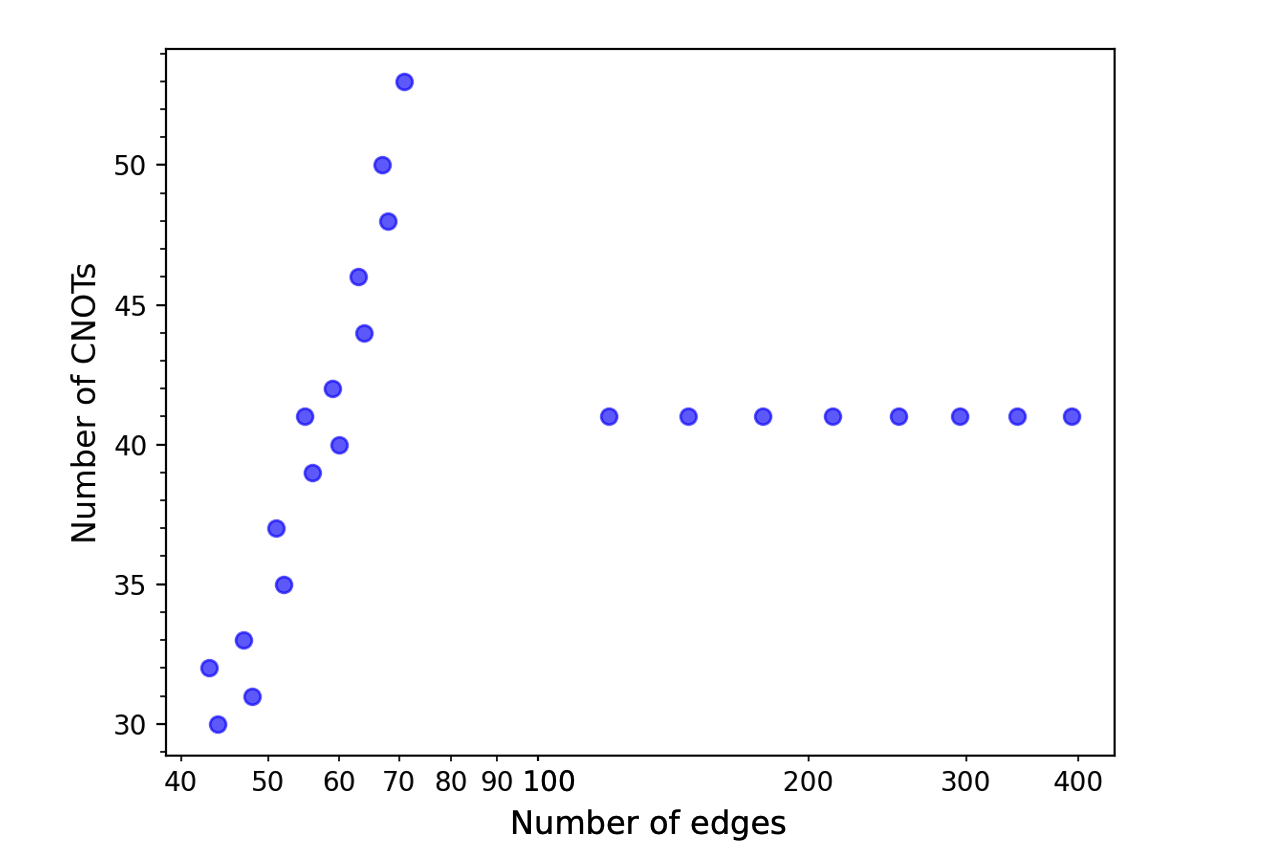}
    \caption{Number of \cnot{}s vs Number of edges (x-axis in logarithmic scale) correlation diagram of a 30-qubit RGS. Points correspond to graphs in the orbit. The orbit shows two distinct behaviors, a group of members showing a linear correlation while the others are constant in the number of \cnot{}s.}
    \label{fig:kaur_correlation}
\end{figure}

\section*{Appendix B: Repeater graph performance}
\label{App:B}
A repeater graph is made up of a set of core nodes forming a complete sub-graph, and a set of leaf nodes (arms) attached to each one of the core qubits. The final goal of a repeater state can be thought of as distilling a perfect maximally entangled EPR pair that is established between two distant parties by distributing the leaf qubits (\Cref{fig:rgs2epr}). Therefore, the overall performance of a single repeater graph state can be determined by evaluating the quality of this distilled 2-qubit state that can be quantified, for instance, via its fidelity with a maximally entangled state.

\begin{figure*}
    \centering
    \includegraphics[width=0.8\textwidth]{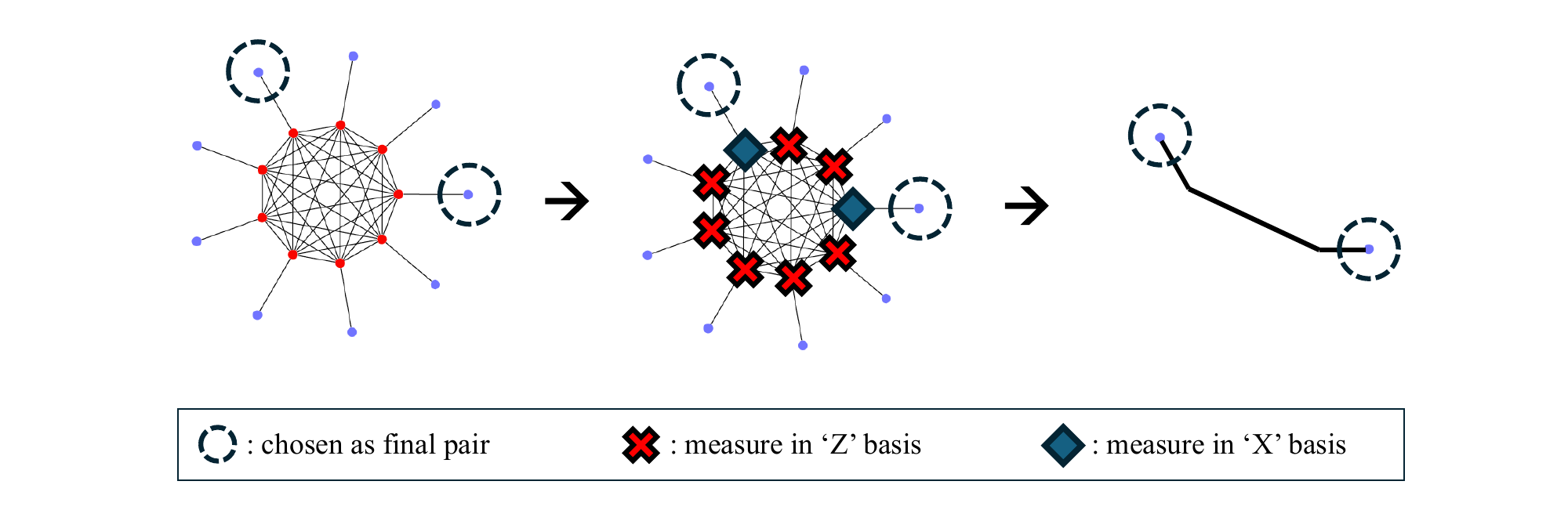}
    \caption{EPR distillation process. From left to right, first a pair of qubits is selected to comprise the final two-qubit state, the rest of the qubits, except for the two nodes connected to the chosen pair are measured in the $Z$ basis, and finally the two neighboring nodes are measured in the $X$ basis. The resultant state is locally equivalent to a graph where the two qubits are connected to each other directly, representing a maximally entangled state.}
    \label{fig:rgs2epr}
\end{figure*}

We now estimate this final fidelity for two cases, a 10-qubit and a 20-qubit RGS, and for each case we consider 2 scenarios: (i) using the original generation circuit and (ii) using the optimized one with reduced \cnot{}s. Considering the emitter-emitter \cnot{} gates to be the dominant source of imperfection, we applied a noise model with an added depolarizing channel (with a depolarization rate of 1\%) to each of the \cnot{} gates in the generation circuit. For each case, we simulate the distillation procedure for different choices of the two arms used to form the EPR pair, and report the lower-bound for the fidelity of the final state. The results are listed in the table \Cref{tab:epr_fids} and indicate that the same fidelity for the non-optimized scenario can be achieved by using the optimized circuit for an RGS twice the size. 
\begin{table}[h]

\begin{minipage}{\columnwidth}
\renewcommand{\arraystretch}{1}
\setlength{\tabcolsep}{20pt} 
\centering
\caption{Fidelity of the final entangled pair distilled from a repeater graph state}
\resizebox{\linewidth}{!}{
\begin{tabular}{|c||c|c|}
\hline
\multicolumn{1}{|l||}{ \hspace{2.5mm} RGS size} & non-optimized & optimized\\
\hline
\ n=10 & \colorbox{yellow!40}{$88\pm1\%$}  & $95\pm1\% $  \\
\hline
\ n=20 & $75\pm1\%$ & \colorbox{yellow!40}{$88\pm1\%$}   \\
\hline
\end{tabular}
}%
\label{tab:epr_fids}
\end{minipage}
\end{table}
Simulating for a range of sizes and depolarization rates, the results remain consistent, confirming that the optimization allows for doubling the size of the generated RGS state without affecting the performance of the repeater. 

Such an increase in size can lead to orders of magnitude increase in the achievable quantum key distribution rate. Specifically, the repeater protocol's success rate is given by $(1-(1-{P_b}^{n/4}))^N$ \cite{azuma_all-photonic_2015}. Here, $n$ is the number of photons in the RGS, $N$ is the number of repeater stations between the two parties, and $P$ is the probability of a successful Bell state measurement between two photons from adjacent repeater stations. Using the parameters from \cite{azuma_all-photonic_2015}'s realistic example cases, and considering a total distance of $400$ $km$, with $N=100$ and $P_b=0.416$, we find a $2.66\times10^5$ increase in success rate when $n$ is doubled from $15$ to $30$ qubits. The value for $P_b$ was obtained by multiplying the probability of both photons making it to the Bell state measurement (BSM) station, by the success probability of the BSM itself ($= 50\%$ \cite{browne_resource-efficient_2005}). Assuming a loss of $0.2$ $dB/km$ and $2$ $km$ travel distance for each photon from two adjacent RGS to reach the BSM station in the middle (half the total distance $=400/100=4$ $km$ between two repeaters) the probability of having both photons present is equal to $0.832$ and so $P_b = 0.5\times0.832 = 0.416$.

\section*{Appendix C: Correlation analysis}
\label{App:C}
In a correlation analysis of the local equivalency class of a graph, the goal is to determine whether there exists a significant correlation between a graph metric (a parameter related to the shape of a graph) and a cost parameter of its generating circuit. We examine how changes in the average values of one metric (chosen as the dependent variable) correspond to variations in the other metric (chosen as the independent variable) across the locally equivalent graphs in an orbit or in a sample from it. For instance, in \Cref{fig:n_edge_n_unitaries}, the independent variable (x-axis) is set to be the number of edges in the graph and the dependent variable (y-axis) is chosen as the average of the circuit's gate-count over the graphs sharing the same number of edges in the sample taken from the orbit. This diagram is derived for a sample of the LC orbit of a random graph of size $n$=20 with an edge probability of $p$=0.95. The Pearson correlation coefficient, used to measure the degree of linear relationship between two variables \cite{Pearson}, for this specific diagram is $94\%$.
Repeating the same procedure for 100 random graphs of the same type ($n$=20, $p$=0.95), we find the distribution of the correlation coefficients, as seen in \Cref{fig:histogram}. This shows an average Pearson correlation coefficient of $0.86\pm0.07$, indicating that for most of the graphs of this type, a strong correlation—Pearson coefficient over 0.5—holds true. 

\begin{figure}[h]
\centering%
\includegraphics[width=0.9\columnwidth]{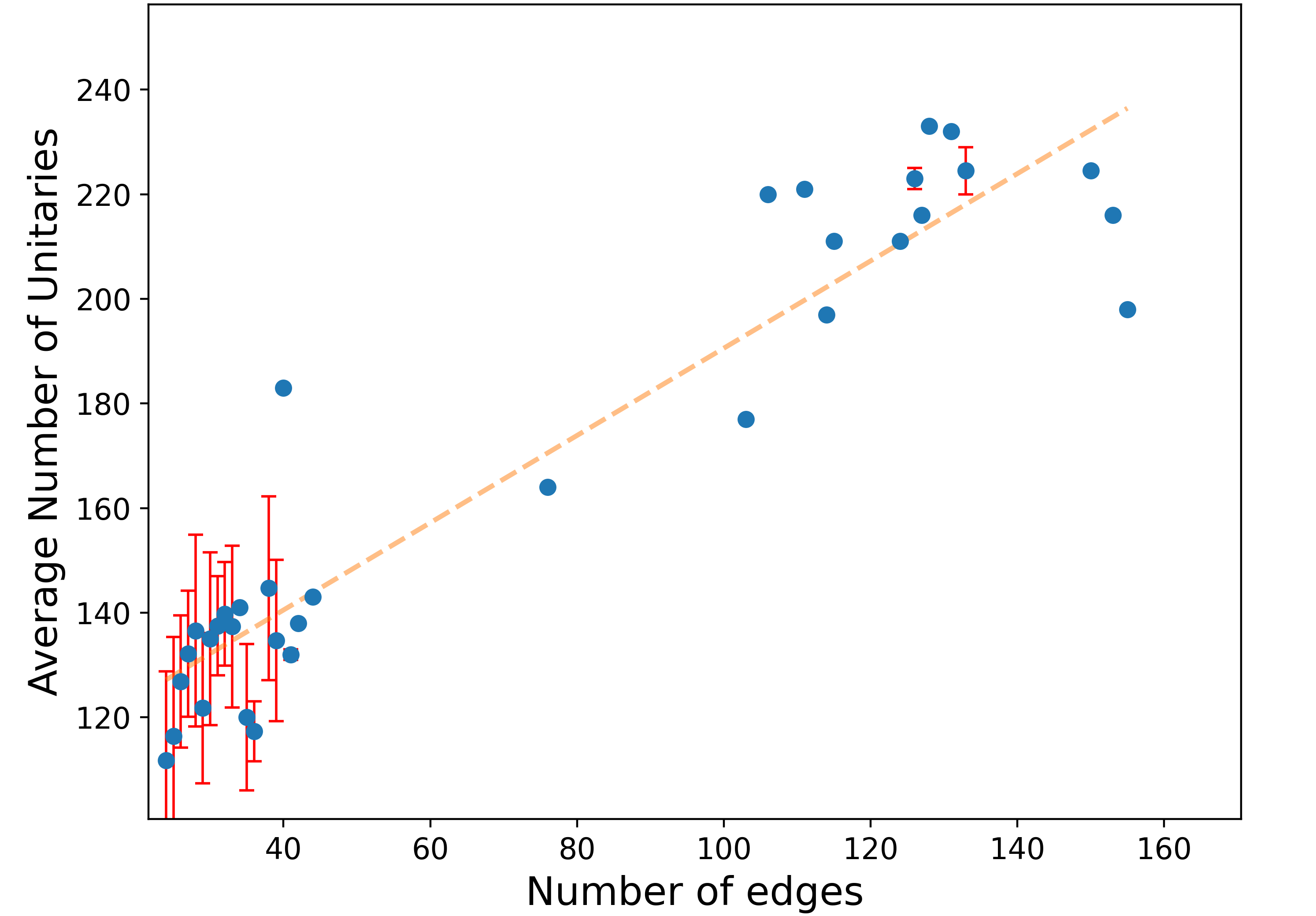}
\caption{The correlation diagram for the average number of unitary gates in a circuit vs the number of edges, over a sample of size 200 from the orbit of the graph. The error bars indicate one standard deviation when more than one graph is associated with the same number of edges in the sample.}
\label{fig:n_edge_n_unitaries}
\end{figure}

\begin{figure}[h]
\centering%
\includegraphics[width=\columnwidth]{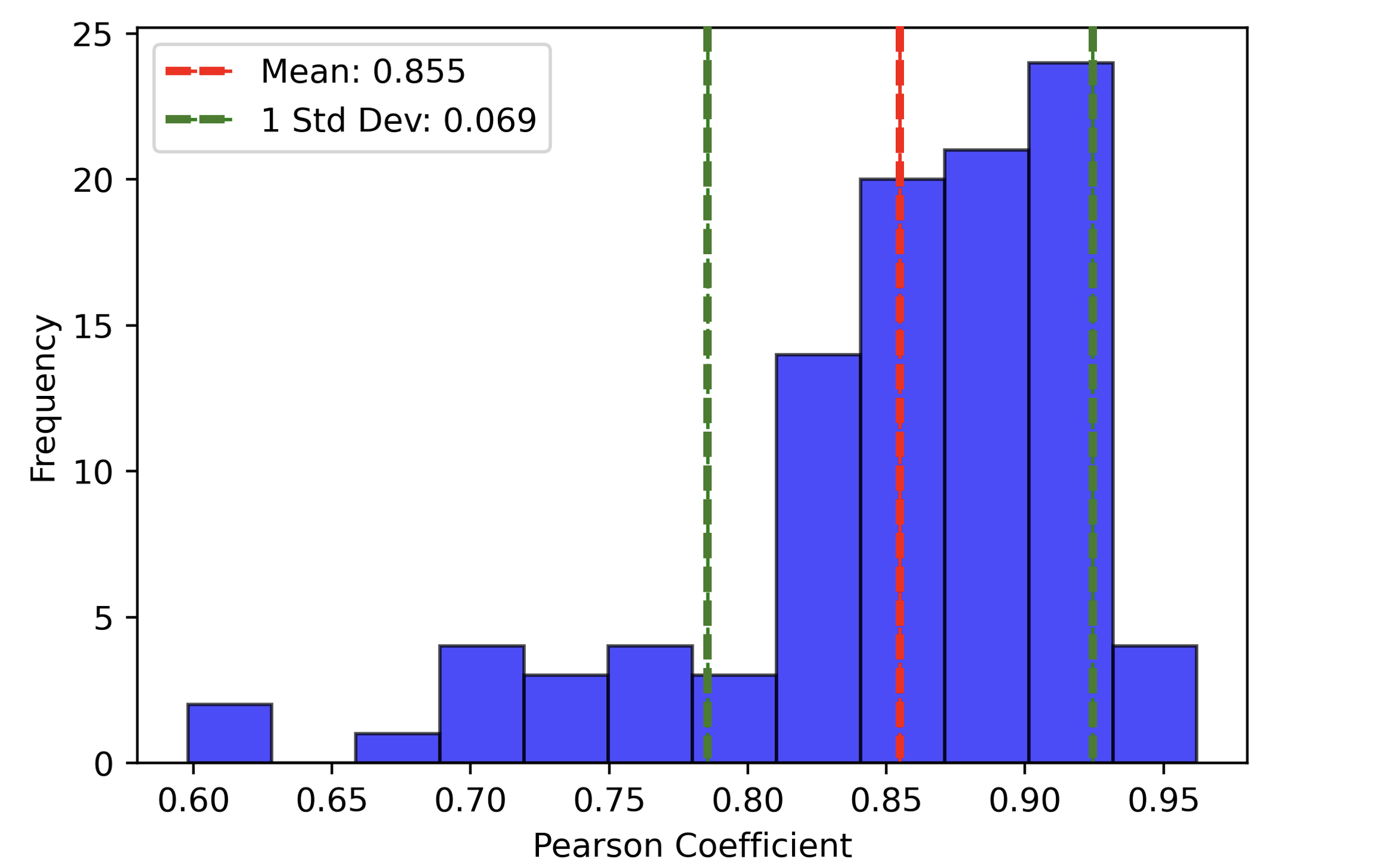}
\caption{The distribution of Pearson coefficients for `edge count vs gate count' correlations for 100 trials with different random graphs of size 20 and edge probability 0.95.}
\label{fig:histogram}
\end{figure}

With the correlation confirmed, one can, in this case, aim for the graphs in the orbit that have a smaller number of edges, corresponding to circuits with fewer unitary gates overall. To do this, we developed an active edge reduction strategy to get to the sparser graphs in the orbit through applying local complementation operations on specific nodes, starting from the initial graph. In particular, we select the node with the highest product of node degree (number of neighbors) and clustering coefficient \cite{watts_collective_1998}, and apply local complementation on it. The clustering coefficient of a node is defined as the ratio of the number of existing edges in the neighborhood of that node, to the maximum possible number of edges in that neighborhood. The process is iteratively repeated on the resulting graph until further steps would no longer reduce the number of edges.

This active edge reduction strategy shows a one to two order of magnitude optimization speed up compared with a random search method, where a random sample from the graphs in the orbit is taken, converted into corresponding circuits, and the best circuit (with minimum unitary gate count) is selected. \Cref{fig:runtimes} shows a runtime scaling diagram for graphs of size n=30 to 55 while the sample size is chosen so that the performance of both methods is similar for each graph size. This can be seen in \Cref{fig:performance}.  

The edge reduction strategy used here takes at most $\mathcal{O}(n^2)$ steps to complete. This is because in each step we require the number of edges to be reduced at least by one, and there are at most $n(n-1)/2$ edges in a graph. The clustering coefficient of the graph needs to be calculated in each step, which has a time complexity of $\mathcal{O}(n^3)$ using a brute-force algorithm. Therefore, the total complexity will be upper-bounded by $\mathcal{O}(n^5)$. However, in practice, the actual scaling with size is much smaller than the worst case scenario, for example, in the specific use case of dense graphs (edge probability 95\%), the scaling turns out to be $n^{2.68}$ for our python implementation (see \Cref{fig:strategy_scaling}). The method may thus be used for optimization of graphs with thousands of nodes within minutes on commercial processors.

\begin{figure}[h]
\centering%
\includegraphics[width=0.9\columnwidth]{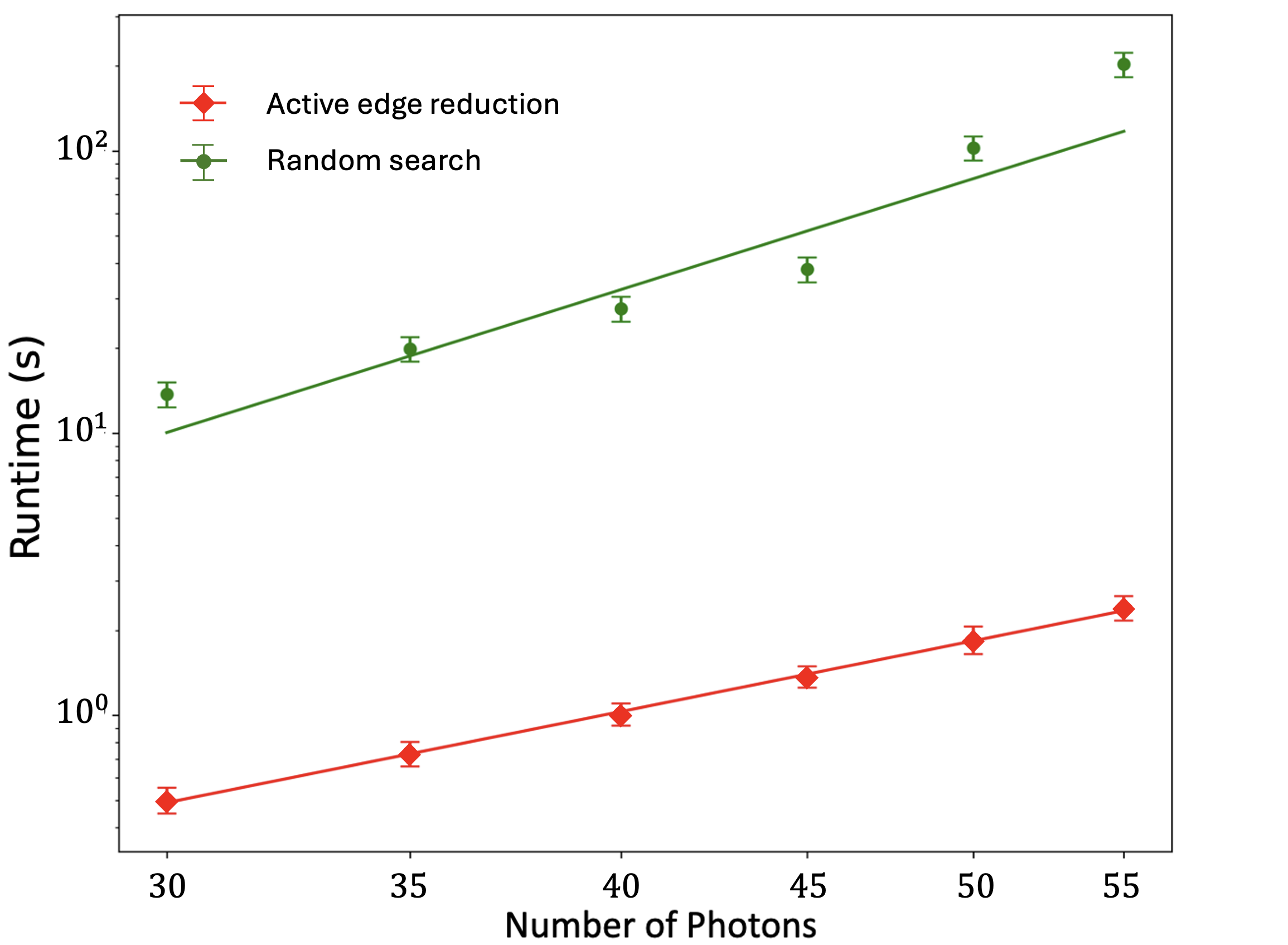}
\caption{Runtime comparison for the same performance between the active edge reduction method and a random search method. Error bars are 1 standard deviation for data averaged over 100 trials.}
\label{fig:runtimes}
\end{figure}

\begin{figure}[h]
\centering%
\includegraphics[width=0.95\columnwidth]{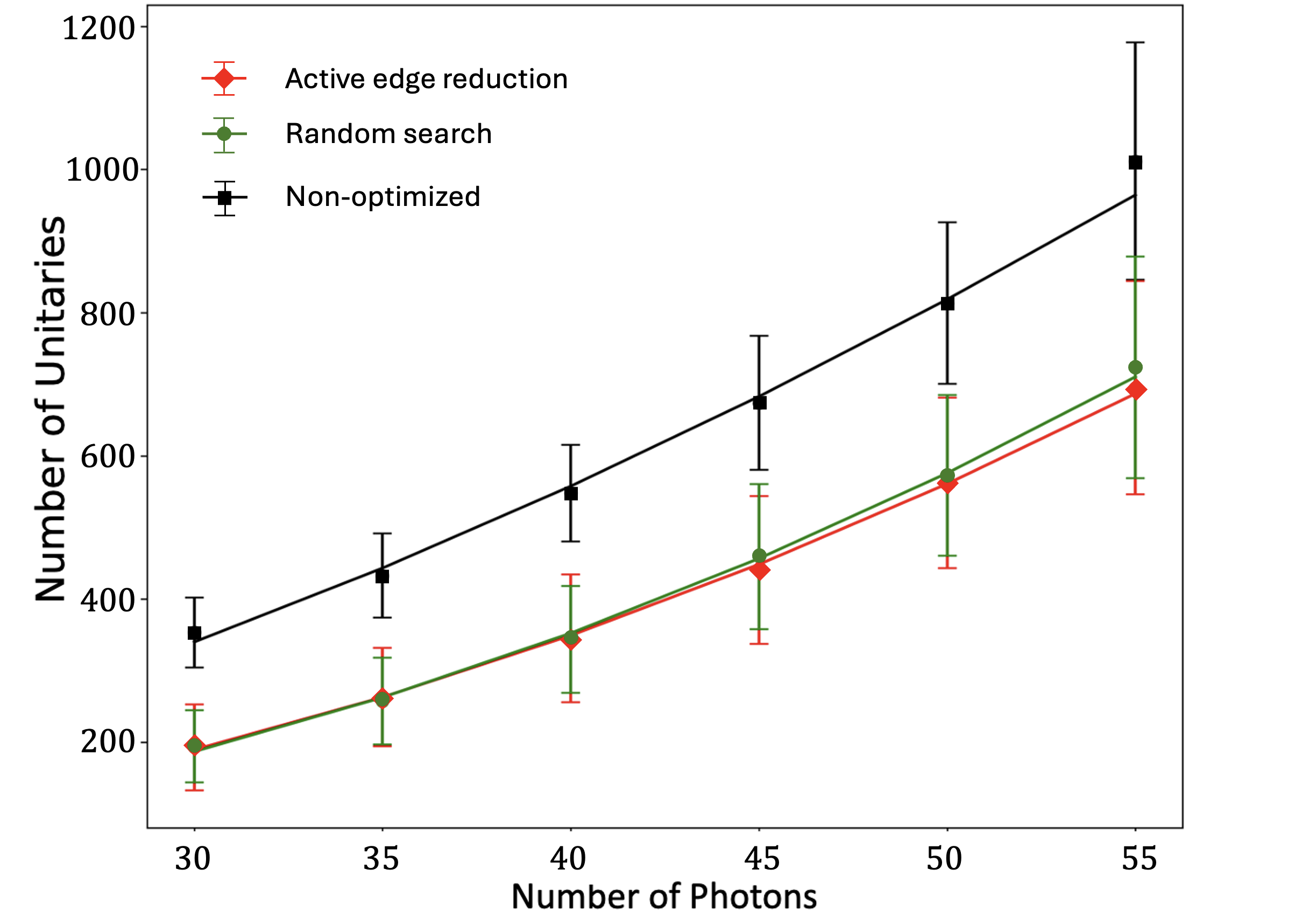}
\caption{Unitary gate count for optimization with the active edge reduction strategy (red diamond), and the random search method (green circle). The sample size for random search is chosen such that the performance would be similar to the active method. The initial non-optimized gate count (black square) is also depicted for comparison. Error bars represent one standard deviation for each data point averaged over 100 trials.}
\label{fig:performance}
\end{figure}

\begin{figure}[h]
\centering%
\includegraphics[width=0.9\columnwidth]{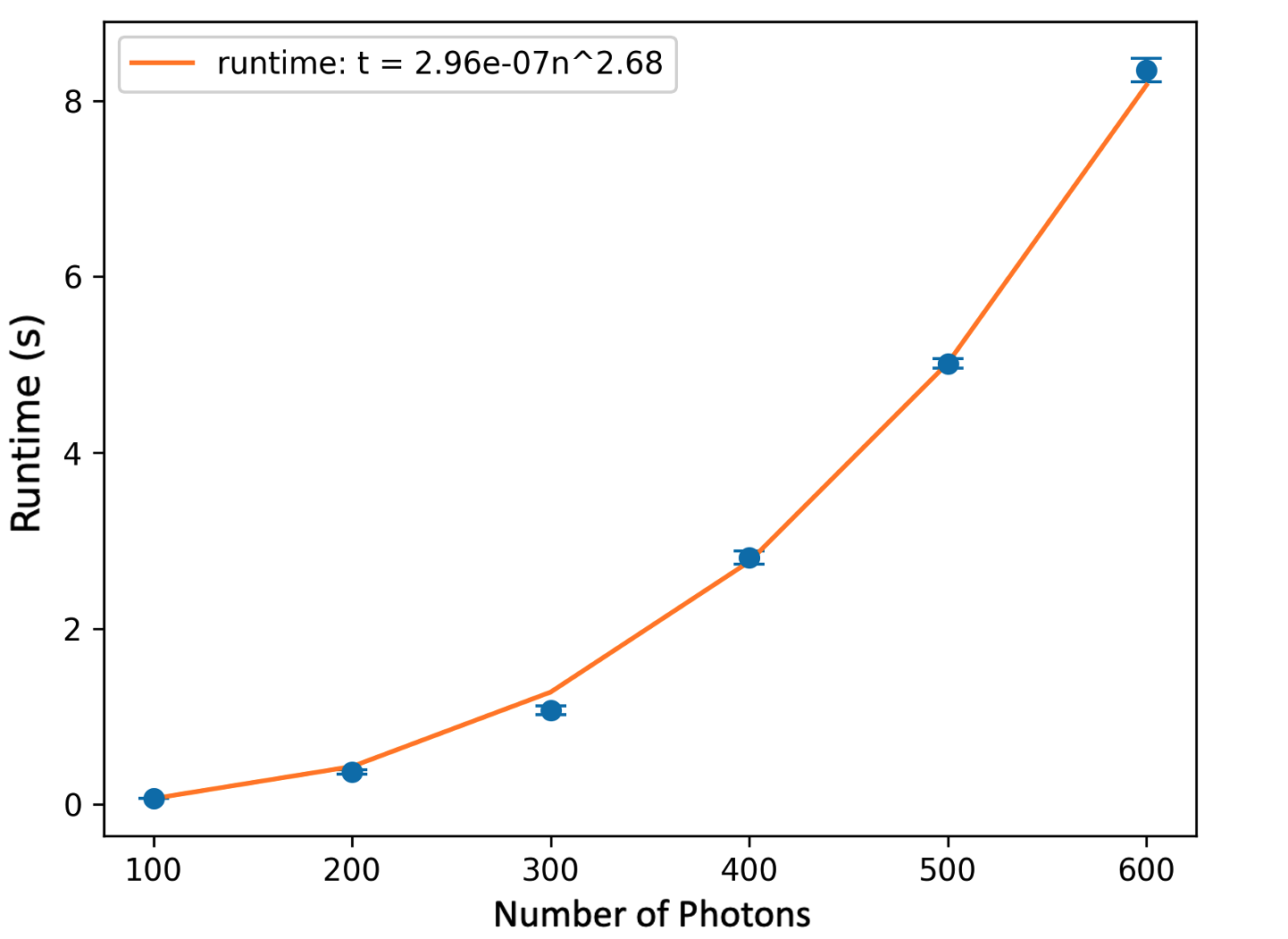}
\caption{Polynomial scaling of the runtime with graph size for the proposed edge reduction strategy is evident from the line of best fit. Error bars represent one standard deviation for each data point averaged over 100 trials.}
\label{fig:strategy_scaling}
\end{figure}

\section*{Appendix D: Exhaustive analysis of small graphs}
\label{App:D}

To showcase the results of the exhaustive analysis, we look into an entanglement class associated with the graph (A) in \Cref{fig:class17}, i.e., all graph states that are LC equivalent to this graph up to relabeling of the nodes. We have identified multiple intermediate target graphs to utilize in the generation process of any member of this entanglement class, each tailored for optimizing a certain cost parameter. In \Cref{fig:class17}, examples of a case with the fewest number of \cnot{}s (B), a case with the smallest emitter depth (C), and a case with minimal number of unitary gates (D) are presented. More details for each option are listed in \Cref{tab:class17}. We remark that there might be more than one best performing intermediate graphs depending on the cost function used. One can find the maximum improvement for each cost parameter by comparing the best values with the worst ones present in the last column, showing the maximum of the cost parameters over this class when no optimization is done. Overall, any arbitrary cost function can be employed to filter the results of such an exhaustive analysis, e.g., the fidelity of the final state with the target when a proper noise model is employed. The full dataset containing the results from the analysis of all 45 entanglement classes of graphs with $n\leq7$ nodes is available at Ref. \cite{dataset}. 

\begin{figure}[h]
\centering%
\includegraphics[width=\columnwidth]{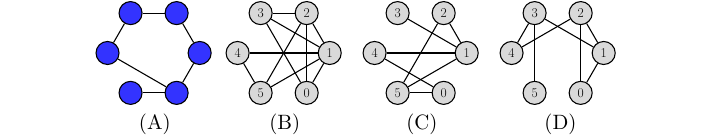}
\caption{A set of optimal intermediate graphs (B-D) for the generation of the entanglement class associated with graph (A), minimizing different cost functions in the following order from (B) to (D): \cnot{} count, emitter depth, and unitary gate count. There might be more than one best graph for each metric in the orbit. Node labels indicate the photon emission order.}
\label{fig:class17}
\end{figure}
\begin{table}[]

\begin{minipage}{\columnwidth}
\renewcommand{\arraystretch}{1.3}
\setlength{\tabcolsep}{10pt} 
\centering
\caption{Summary of the cost metrics for the alternative graphs outlined in \Cref{fig:class17}.}
\resizebox{\linewidth}{!}{
\begin{tabular}{|c||c|c|c|c|c|}
\hline
\multicolumn{1}{|l||}{ \hspace{2.5mm} Metrics} & Graph $A$ & Graph $B$ & Graph $C$ & Graph $D$ & Orbit Max\\
\hline
\# Emitters & $2$  & $2 $ & $3$ & $   2  $ & $   3   $ \\
\hline
\# \cnot{}s & 3 & 2  & 3 & 2& 10    \\
\hline
Depth\footnotemark[1] & 14 & 23 &7 & 10& 41\\
\hline
\# Unitaries & 19 & 48 & 18 & 16& 96 \\
\hline
\end{tabular}
}%
\footnotetext[1]{Emitter depth as defined in \Cref{background}}
\label{tab:class17}
\end{minipage}
\end{table}

\bibliography{Refs.bib}
\end{document}